\documentstyle[sprocl]{article} 
\input{psfig} 
\begin{document} 
\title       {SOLVING GAUGE-FIELD THEORY \\ BY 
                 DISCRETIZED LIGHT-CONE QUANTIZATION
\footnote{
Talk at the International Workshop
``Perspectives of strong coupling gauge theories (SCGT96)'',
Nagoya (Japan), November 13-16, 1996;
  to appear in the proceedings.}}
\author{H.C. PAULI}
\address{Max-Planck-Institut f\"ur Kernphysik \\
                   D-69029 Heidelberg}

\maketitle\abstracts{ 
The canonical front form Hamiltonian for non-Abelian  SU(N)
gauge theory in 3+1 dimensions is mapped
non-perturbatively 
on an effective  Hamiltonian which acts 
only in the Fock space of a quark and an antiquark.
The approach is based on the novel method of iterated
resolvents and on discretized light-cone quantization, 
driven to the continuum limit. It is free of the usual
Tamm-Dancoff truncations of the Fock space, rather
the perturbative series are consistently resumed to all orders
in the coupling constant. 
Emphasis is put on  dealing with the many-body
aspects of gauge field theory. The effective interaction 
turns out to be the kernel of an integral equation
in the momentum space of a single quark, which
is frame-independent and solvable on
comparatively small computers. 
Important is that the higher Fock-space amplitudes 
can be retrieved self-consistently  from these solutions.
} 
 
\section{The Hamiltonian matrix}
\label{sec:2} 

In canonical field theory the  four  components of
the  energy-momentum vector $ P ^\nu$ commute and 
are constants of the motion. 
In the front form of Hamiltonian dynamics \cite{dir49} 
they are denoted \cite{leb80} by 
$ P ^\nu = (P  ^+, \vec P  _{\!\bot}, P  ^-)$.
Its spatial components $\vec P  _{\!\bot}$  and $P  ^+$ 
are independent of the interaction and diagonal in
momentum representation.
Their eigenvalues are the sums of the single particle
momenta, 
$     P ^+ = \sum p^+ = {2\pi \over L } K$ and
$     \vec P  _{\!\bot} = \sum\vec p _{\!\bot}$. 
Each single particle has four-momentum 
$ p^\mu = (p^+, \vec p _{\!\bot}, p^- )$ and 
sits on its mass-shell $ p^\mu p_\mu = m^2$. 
Each particle state ``$q$''  is then characterized by six 
quantum numbers
$     q = (p^+, \vec p_{\!\bot}, \lambda;c, f ) $: 
the three spatial momenta, helicity, color and flavor.
The temporal component $P ^-= 2 P_+ $ depends on the
interaction  and is a very complicated non-diagonal
operator \cite{brp91}. It propagates the system in the 
light-cone time $ x^+ = x^0 + x^3$.
The contraction of  $P ^\mu$  is the operator of
invariant mass-squared,
\begin {equation} 
      P ^\mu P _\mu  = P ^+ P ^- - \vec P  _{\!\bot} ^{\,2} 
      \qquad  \equiv H _{\rm LC} \equiv H
\ .\end {equation} 
It is Lorentz invariant and  referred to \cite{brp91}
somewhat improperly but conveniently 
as the `light-cone Hamiltonian' 
$ H _{\rm LC} $, or shortly $ H $. 
One seeks a representation in which $H$ is diagonal
\cite{brp91,pab85a}, 
$ H \vert \Psi\rangle = E\vert \Psi\rangle $.
Introductory texts \cite{brp91,gla95} are available. 

By convenience, the (light-cone) Hamiltonian is split into 
four terms:
$      H = T + V + F + S$.
The kinetic energy  $T$ survives the limit of the coupling constant 
$ g $ going to zero. Since it is diagonal in Fock-space representation, 
its eigenvalue is the {free invariant mass squared}
of the particular Fock state. 
The vertex interaction $ V $ is the relativistic interaction
{\it per se}. It is linear in $g$ and changes the particle number 
by 1. Those matrix elements of $V$ which
change the  particle number by 3 are present in the instant form,
but vanish for light-cone kinematics:  
The vacuum {\em does not fluctuate}.
The instantaneous interactions $ F $ and $ S $ are 
consequences of working in the light-cone gauge  $A^+=0$.
They are proportional to $g^2$.  
By definition, the seagull interaction $S$  conserves
the particle number and the fork interaction $F$ changes
it by 2. 
\begin{figure} 
\psfig{figure=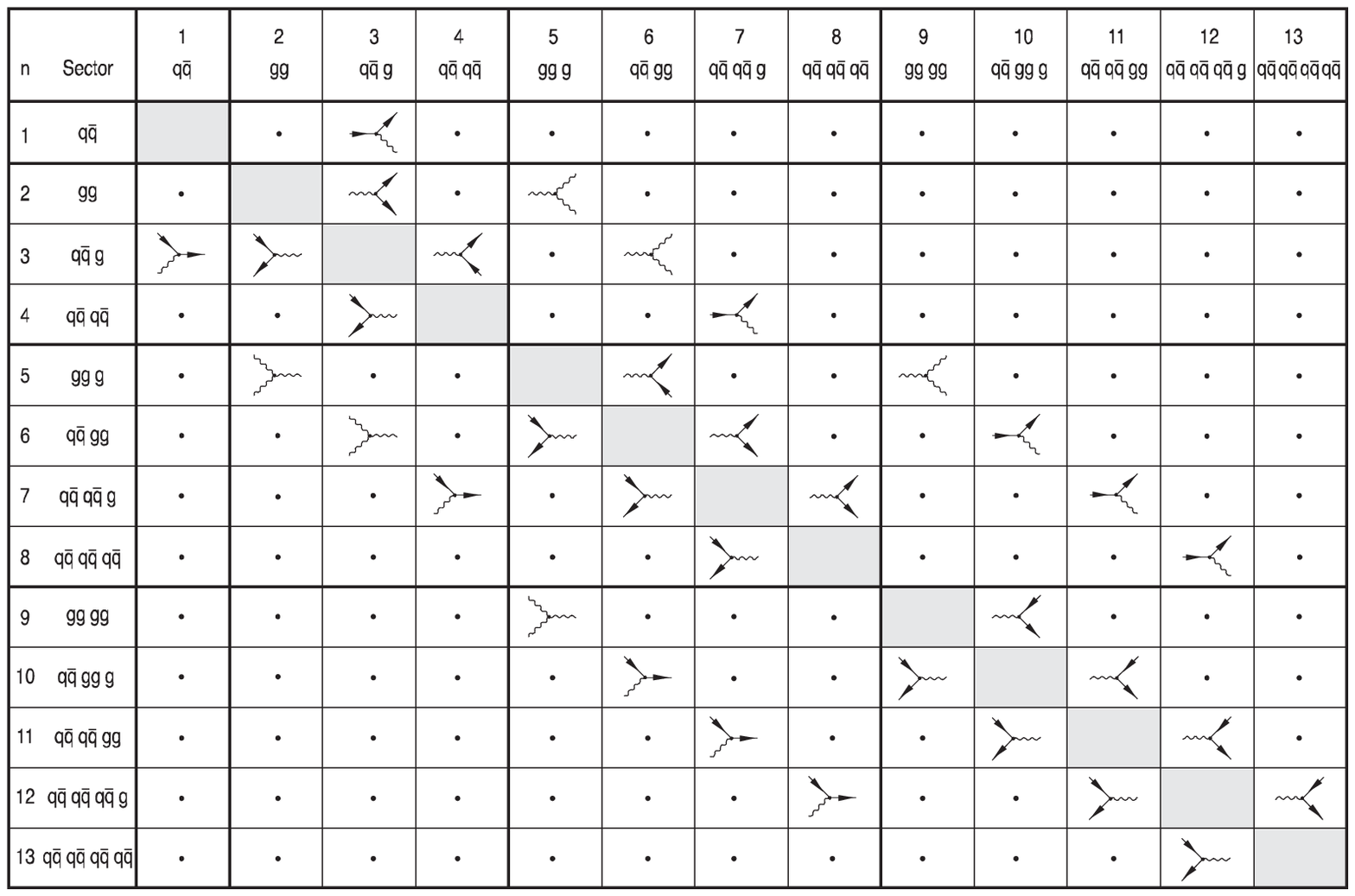,width=120mm} 
\caption{\label{fig:holy-2} 
    The Hamiltonian matrix for a SU(N)-meson and 
     a harmonic resolution $K=4$.
     Only vertex diagrams are included.
     Zero matrices are marked by ($\cdot$). 
} \end{figure} 

The eigenfunctions
$\vert\Psi\rangle$ are complicated
superpositions  of Fock-space projections with 
contributions like $\langle q\bar q\vert\Psi\rangle$ 
or $\langle q\bar q\,g\vert\Psi\rangle$. 
The eigenvalue equation stands thus for an infinite set of 
coupled integral equations which are even difficult to
write down. However, if 
one works with periodic  boundary conditions, 
with discretized light-cone quantization \cite{pab85a} (DLCQ), 
the infinite set of coupled integral equations is mapped onto a
{\em finite set} of coupled matrix equations, as illustrated in 
Fig.~\ref{fig:holy-2}.
As consequence of discretization, the Fock states are 
denumerable and orthonormal.
The {momentum fraction} carried by the  $n$-th particle
is $ x_n  = p^+_n   /P ^+ $. The sum of all momentum fractions
and the sum of all transversal momenta is constrained to
$     \sum _n x _n = 1$ and 
$     \sum _n \vec k_{n_\perp} = 0 $,
with the intrinsic transversal momenta $\vec k_{\!\perp}$
defined by  
$\vec p_{n_\perp} = x _n\vec P_{\!\perp}+\vec k_{n_\perp}$.
Since $P  ^+$ has only positive eigenvalues and  
since each particle has a lowest possible value of $p^+$, 
the number of particles in a Fock state is limited for any 
fixed value of the {\it harmonic resolution} \cite{pab85a} $ K $. 

Here then is the problem, the bottle neck of any Hamiltonian
approach in  field theory: The dimension of the Hamiltonian
matrix  increases exponentially fast, even if one regulates
the transversal momenta by some suitable cut-off $\Lambda$, 
like the invariant-mass cut-off of Lepage and Brodsky
\cite{leb80} or similar constructs.
Suppose, the regularization procedure allows for 10 discrete
momentum states in each direction. A single particle has then 
about $10^3$ degrees of freedom.
A Fock-space sector with $n$ particles
has then roughly  $10^{n-1}$ different Fock states.
Sector 13 with its  8 particles has thus about  $10^{21}$
and sector 1 ($q\bar q$) about $10^3$ Fock states. 
Deriving an effective interaction can be understood as 
reducing the dimension in a matrix diagonalization problem 
from $10^{21}$ to $10^{3}$!

\section {The effective interaction} 
\label {sec:3}

Effective interactions are a well known tool in  many-body 
physics. In field theory the method \cite{tam45,dan50} is 
known as the Tamm-Dancoff-approach. 
The Fock-space sectors $\vert i \rangle$ appear in a most  
natural way and the Hamiltonian matrix can be understood 
as a matrix of block matrices,
whose rows and columns are enumerated by
$i=1,2,\dots N$ like in Fig.~\ref{fig:holy-2}.
The eigenvalue equation can then be written 
as a coupled block matrix equation:
\begin {equation} 
      \sum _{j=1} ^{N} 
      \ \langle i \vert H \vert j \rangle 
      \ \langle i \vert \Psi\rangle 
      = E\ \langle n \vert \Psi\rangle 
\, \qquad {\rm for\ all\ } i = 1,2,\dots,N 
\ .\label {eq:319}\end {equation} 
The rows and columns can always be split into the $ P $-space  
and the rest, the $Q$-space. The block matrix equation becomes
\begin {eqnarray} 
   \langle P \vert H \vert P \rangle\ \langle P \vert\Psi\rangle 
 + \langle P \vert H \vert Q \rangle\ \langle Q \vert\Psi\rangle 
 &=& E \:\langle P \vert \Psi \rangle 
 \ ,  \label{eq:321} \\   {\rm and} \qquad
   \langle Q \vert H \vert P \rangle\ \langle P \vert\Psi\rangle 
 + \langle Q \vert H \vert Q \rangle\ \langle Q \vert\Psi\rangle 
 &=& E \:\langle Q \vert \Psi \rangle 
\ . \label{eq:322}\end {eqnarray}
Rewrite the second equation as
$      \langle Q \vert E  -  H \vert Q \rangle 
    \ \langle Q \vert \Psi \rangle 
  =   \langle Q \vert H \vert P \rangle 
    \ \langle P \vert\Psi\rangle $,
and observe that the quadratic matrix 
$ \langle Q\vert E -  H \vert Q \rangle $ could be inverted 
to express the Q-space wave-function 
$\langle Q \vert\Psi\rangle $
in terms of $\langle P \vert\Psi\rangle$. 
The eigenvalue $ E $ is unknown at this point.  One introduces
therefore a redundant parameter $\omega$, and 
defines $G _ Q (\omega)   =  \left[
    \langle Q \vert\omega- H \vert Q \rangle\right]^{-1} $.
The so obtained effective interaction
\begin{equation} 
      \langle P \vert H _{\rm eff} (\omega) \vert P \rangle =
      \langle P \vert H \vert P \rangle +
      \langle P \vert H \vert Q \rangle 
      \ G _ Q (\omega) \ %
      \langle Q \vert H \vert P \rangle 
\ .  \label{eq:340}\end{equation} 
acts only in the $P$-space:
$    H _{\rm eff} (\omega ) \vert\Phi_k(\omega)\rangle =
      E _k (\omega )\,\vert\Phi _k (\omega ) \rangle $.
Varying $\omega$ one generates a set of  
{energy functions} $ E _k(\omega) $. 
Whenever one finds a solution to the 
fix-point equation 
$E _k (\omega ) = \omega $, 
one has found one of the true eigenvalues and eigenfunctions. 
The advantage of working with an effective interaction is 
of analytical nature to the extent that resolvents can be
approximated systematically.  The two resolvents
\begin {equation}
     G _Q (\omega) =  {1\over \langle Q \vert 
           \omega - T - U  \vert Q \rangle} 
     \quad{\rm and}\quad 
     G _0  (\omega) =  {1\over \langle Q \vert 
           \omega - T \vert Q \rangle} 
\ , \label {eq:352} \end {equation} 
defined once with and once without the non-diagonal 
interaction in the Hamiltonian $H = T + U$, respectively,
are identically related by 
$     G _Q (\omega) =
     G _0  (\omega) + G_0 (\omega)  \ U \ G _Q (\omega)  $, 
or by the infinite series of perturbation theory
$    G _Q (\omega) =
    G _0(\omega) + G_0 (\omega) U G _0 (\omega) 
    +    G_0 (\omega) U G_0 (\omega) U G _0 (\omega) 
    + \dots\ $.
The point is, of course, that the kinetic energy $T$ is
a diagonal operator which can be trivially inverted to get 
the unperturbed resolvent $G_0 (\omega)$.

Albeit exact in principle, the Tamm-Dancoff-approach (TDA) 
suffers in practice from two serious defects: (1) The approach
is technically useful only, if one truncates the perturbative 
series to its very first term. This destroys
Lorentz and gauge invariance. (2) If one identifies 
$\omega$ with the eigenvalue, as one should, the effective 
interaction has a non-integrable  singularity, 
both in the original instant form \cite{tam45,dan50} 
and in the front form \cite{kpw92,trp96}.

The above can be understood as to reduce the block matrix
dimension  from 2 to 1. But there is no need to identify 
the $P$-space with the lowest sector. 
One also can choose the $Q$-space 
identical with last sector and the $P$-space 
with the rest, $P=1-Q$: 
The same steps as above reduce then the block matrix dimension 
from $N$ to $N-1$. The effective interaction acts in the now 
smaller space. This procedure can of course be iterated, 
but then one has to deal with `resolvents of resolvents', 
or with iterated resolvents \cite{pau96}. 
Ultimately, one arrives at block matrix dimension 1 
where the procedure stops: 
The effective interaction in the Fock-space
sector with only one quark and one antiquark is defined
unambiguously.

Suppose, in the course of this reduction, one has 
arrived at block matrix dimension $n$, 
with $1\leq n\leq N$.  Denote the corresponding effective interaction  
$H_n (\omega)$. The eigenvalue problem reads then
\begin {equation} 
   \sum _{j=1} ^{n} \langle i \vert H _n (\omega)\vert j \rangle 
                      \langle j \vert\Psi  (\omega)\rangle 
   =  E (\omega)\ \langle i \vert\Psi (\omega)\rangle 
\ , \ \quad{\rm for}\ i=1,2,\dots,n
\,. \label {eq:406} \end {equation}
Observe that $i$ and $j$ refer here to sector numbers.  
Now,  like in the above, define the resolvent of the 
sector Hamiltonian $H_n(\omega)$ by
\begin {eqnarray} 
            G _ n (\omega)   
&=&   {1\over \langle n \vert\omega- H_n (\omega)\vert n \rangle} 
\, \label {eq:410} \\  {\rm thus}\qquad  
            \langle n \vert \Psi (\omega)\rangle   
&=&   G _ n (\omega) 
   \sum _{j=1} ^{n-1} \langle n \vert H _n (\omega)\vert j \rangle 
   \ \langle j \vert \Psi (\omega) \rangle 
\,. \label {eq:411} \end {eqnarray}
The effective interaction in the  ($n -1$)-space becomes then 
\begin {equation}  
       H _{n -1} (\omega) =  H _n (\omega)
  +  H _n(\omega) G _ n  (\omega) H _n (\omega)
\label {eq:414} \end {equation}
for every block matrix  element 
$\langle i \vert H _{n-1}(\omega)\vert j \rangle$.  
Everything proceeds like in above,
including the fixed point equation  $ E  (\omega ) = \omega $.
But one has achieved much more: Eq.(\ref{eq:414}) is a 
{\em recursion relation} which holds for all $1<n<N$!
Since one has started from the bare Hamiltonian in the 
last sector, one has to convene that $H_{N}=H$. 
The rest is algebra and interpretation.

\begin{figure} [t]
\begin{minipage}[t]{50mm} 
\makebox[0mm]{}
\psfig{figure=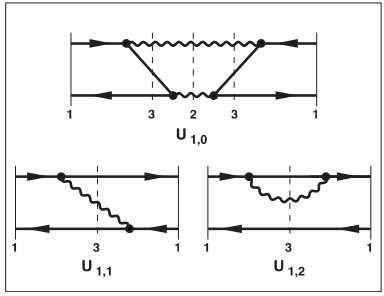,width=50mm} 
\vfill \end{minipage}
\hfill
\begin{minipage}[t]{60mm} \makebox[0mm]{}
\makebox[0mm]{}
\caption{\label{fig:6_1} 
  The three graphs of the effective interaction in the 
  $q\bar q$-space.~---
  The lower two graphs correspond to the chain $U=VG_3V$,
    the upper corresponds to $U_a=VG_3V G_2V G_3V$. 
   Propagator boxes are represented by vertical dashed lines, 
   with the subscript `$n$'  referring to the sector numbers. 
} \vfill \end{minipage}
\end{figure}

Applying the method to the block matrix structure of QCD, 
as displayed in Fig.~\ref{fig:holy-2}, is particularly easy and 
transparent.
By definition, the last sector contains only the diagonal kinetic 
energy, thus  $H_{13}=T_{13}$. Its resolvent is calculated trivially. 
Then $H_{12}$ can be constructed 
unambiguously, followed by $H_{11}$, and so on, until one 
arrives at sector 1. Grouping the so obtained results in a different 
order,  one finds for the sectors with one $q\bar q$-pair: 
\begin {eqnarray} 
     H_{q\bar q} = \qquad     H _1 
&=&  T_1 + V G _3 V + V G _3 V  G _2 V G _3 V 
\ , \label{eq:610}\\  
     H_{q\bar q\,g} = \qquad     H _3 
&=&  T_3 + V G _6 V + V G _6 V  G _5 V G _6 V + V G _4 V 
\ , \label{eq:620}  
\\  
     H_{q\bar q\,gg} = \qquad  H _6 
&=&   T_6 + V G _ {10} V+ V G _{10} V  G _9 V G _ {10} V 
                 + V G _7 V 
\,. \label {eq:622} \end {eqnarray}
The quark-gluon content of  the respective sectors is added here 
for an easier identification. Correspondingly, one obtains  for
the sectors with two $q\bar q $-pairs 
\begin {eqnarray} 
     H_{q\bar q q\bar q} = \qquad     H _4 
&=&  T_4 + V G _7 V + V G _7 V  G _6 V G _7 V  
\ , 
\\  
     H_{q\bar q q\bar q\, g} = \qquad     H _7 
&=&   T_7 + V G _ {11} V+ V G _{11} V  G _{10} V G _ {11} V 
                 + V G _8 V 
\,. 
\end {eqnarray}
In the pure glue sectors, the structure is even simpler:
\begin {eqnarray} 
     H_{gg} = \qquad     H _2 
&=&  T_2 + V G _3 V + V G _5 V 
\ ,\label{eq:643} 
\\  
     H_{gg\, g} = \qquad     H _5 
&=&   T_5 + V G _ {6} V + V G _{9} V 
\ . \label{eq:645}\end {eqnarray}
Note that the above relations are exact. They hold \cite{pau96} 
for an arbitrarily large $K$ and thus in the continuum limit.
Note also that the vertex interaction appears only in even pairs, 
typically in the combination  $VGV$. 
This is the deeper reason why the instantaneous interactions 
can be implemented \cite {pau96} {\it ex post}.
The effective Hamiltonian $ H_{\rm eff} = H_{q\bar q}$ as  
given by Eq.(\ref{eq:610}) 
is illustrated in Fig.~\ref{fig:6_1}. 
As a net result the interaction scatters a quark with helicity
$\lambda_q$ and four-momentum 
$p = (xP^+, x\vec P_{\!\perp} + \vec k_{\!\perp}, p^-)$
into a state with $\lambda_q^\prime$ and four-momentum 
${p^\prime} = (x^\prime P^+, x^\prime\vec P_{\!\perp} 
+ \vec k_{\!\perp}^\prime, {p^\prime}^-)$.

In the {continuum limit}, the resolvents are replaced by 
propagators and the eigenvalue problem 
$H _{\rm eff}\vert \psi\rangle=M^2\vert \psi\rangle$ 
becomes an {integral equation} 
\begin{eqnarray} 
    M_b^2\langle x,\vec k_{\!\perp}; \lambda_{q},
    \lambda_{\bar q}  \vert \psi _b\rangle =
    \left[ {\widetilde m_{q} + \vec k_{\!\perp}^2 \over x } +
    {\widetilde m_{\bar  q} + \vec k_{\!\perp}^2 \over 1-x } \right]
    \langle x,\vec k_{\!\perp}; \lambda_{q},
    \lambda_{\bar q}  \vert \psi _b\rangle &&
\nonumber\\
    +\sum _{ \lambda_q^\prime,\lambda_{\bar q}^\prime}
    \!\int_D\!dx^\prime d^2 \vec k_{\!\perp}^\prime\,
    \langle x,\vec k_{\!\perp}; \lambda_{q}, \lambda_{\bar q}
    \vert U+U_a\vert x^\prime,\vec k_{\!\perp}^\prime; 
    \lambda_{q}^\prime, \lambda_{\bar q}^\prime\rangle\,
    \langle x^\prime,\vec k_{\!\perp}^\prime; 
    \lambda_{q}^\prime,\lambda_{\bar q}^\prime  
    \vert \psi _b\rangle &&
\ .\label{eq:445}\end {eqnarray}
The domain $D$ restricts integration 
in line with regularization.
The effective potential $U$ is diagrammatically defined
in diagram $U_{1,1}$ of Fig.~\ref{fig:6_1} and cannot change 
the flavor of the quark.  It includes all fine and hyperfine interactions. 
Diagram $U_{1,2}$ contributes to the effective quark mass
$\widetilde m$.  The flavor-changing annihilation interaction 
$U_a=U_{1,0}$ is probably of less importance in a first assault.
The eigenvalues $M_b^2$ are the invariant mass$^2$  of a physical 
particle and the corresponding wave-function 
$      \langle x,\vec k_{\!\perp}; \lambda_{q},
        \lambda_{\bar q}  \vert \psi _b\rangle$ 
gives the probability amplitudes for finding 
a flavored quark with momentum fraction $x$, 
intrinsic transverse momentum $\vec k_{\!\perp}$ 
and helicity $\lambda_{q}$. 
These are boost-invariant quantities. 
The eigenfunctions represent the normalized projections 
of the full eigenfunction $\vert\Psi\rangle$ onto the Fock states
$ \vert q;\bar q\rangle 
   = b^\dagger _q d^\dagger_{\bar q} \vert vac \rangle $. 

The knowledge of $\psi_b$ is  sufficient to retrieve all desired 
Fock-space  components of the total wave-function. 
The key is the upwards recursion relation Eq.(\ref{eq:411}).
Obviously, one can express the higher Fock-space components 
$\langle n\vert\Psi\rangle$ as functionals of $\psi_{q\bar q}$ 
by a finite series of quadratures, {\it i.e.} of matrix multiplications 
or of momentum-space integrations.  
One need not solve another eigenvalue problem.  
I show this by calculating the probability amplitude 
for a  $\vert gg\rangle$- or a  
$\vert q\bar q\,g\rangle$-state
in an eigenstate of the full Hamiltonian.
The first two equations of the recursive set in Eq.(\ref{eq:411}) are
\begin {eqnarray} 
        \langle 2 \vert\Psi\rangle &=& 
        G _ 2 \langle2\vert H_2\vert1\rangle 
        \langle 1 \vert \Psi  \rangle 
\,,\label{eq:448}\\  {\rm and}\quad
        \langle 3 \vert\Psi\rangle  &=& 
        G _ 3 \langle 3 \vert H _3 \vert 1 \rangle 
        \langle 1 \vert \Psi  \rangle +
        G _ 3 \langle 3 \vert H _3 \vert 2 \rangle 
        \langle 2 \vert \Psi  \rangle 
\,.\end{eqnarray}
The sector Hamiltonians $H _n$ have to be substituted from
Eqs.(\ref{eq:620}) and (\ref{eq:643}).
In taking block matrix elements of them, the formal expressions
are simplified considerably since many of the Hamiltonian blocks 
in Fig.~\ref{fig:holy-2} are zero.  
One thus gets simply
$\langle2\vert H _2\vert1\rangle=\langle2\vert VG_3V\vert1\rangle$
and therefore
$\langle2\vert\Psi\rangle=G_2VG _3V \langle1\vert\Psi \rangle$.
Substituting this into Eq.(\ref{eq:448}) gives 
$\langle 3 \vert \Psi \rangle  = G _ 3 V\langle 1 \vert\Psi \rangle +
    G _ 3 VG _ 2 VG _3 V \langle1\vert\Psi\rangle$. 
These findings can be summarized more succinctly as
\begin {eqnarray} 
    \vert \psi_{gg} \rangle  &=& 
    G _ {gg} VG _{q\bar q\,g}V\:\vert\psi_{q\bar q}\rangle  , 
\\  {\rm and } \qquad
    \vert\psi_{q\bar q\,g}\rangle  &=& 
    G _ {q\bar q\,g} V\:\vert\psi _{q\bar q}\rangle +
    G _ {q\bar q\,g} VG _ {gg} VG _{q\bar q\,g} V 
    \:\vert\psi_{q\bar q}\rangle 
\ .\end{eqnarray}
Note that the above relations are exact. 
The finite number of terms is in strong contrast to the infinite 
number of terms in perturbative series. Iterated resolvents sum 
the perturbative series to all orders in closed form.

\section{The mean field and the vertex functions} 
\label{sec:5}

\begin{figure} [t]
\begin{minipage}[t]{65mm} 
\makebox[0mm]{}
\psfig{figure=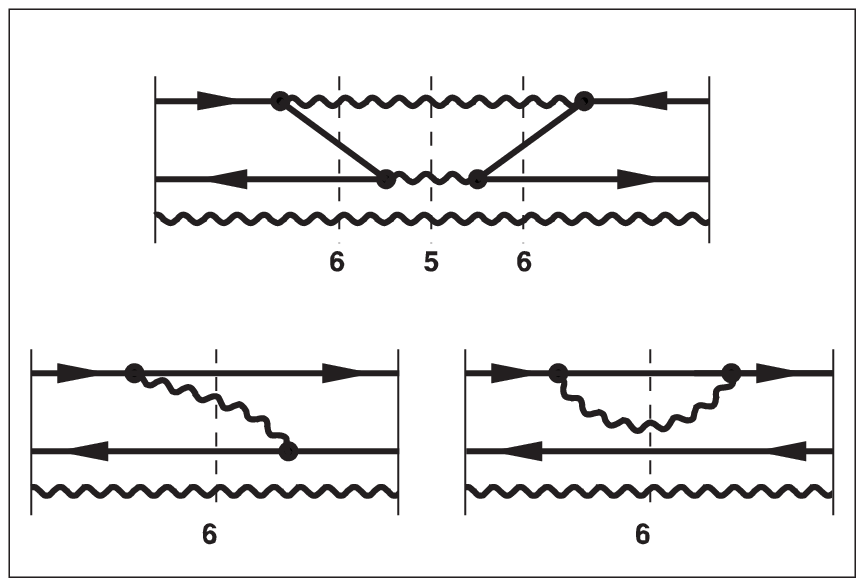,width=67mm} 
\vfill
\caption{\label{fig:6_2} 
    The three graphs of the  spectator interaction 
    in the $q\bar q\,g$-space. 
    Note the role of the gluon as a spectator.
}\end{minipage}
\hfill
\begin{minipage}[t]{50mm} 
\makebox[0mm]{}
\psfig{figure=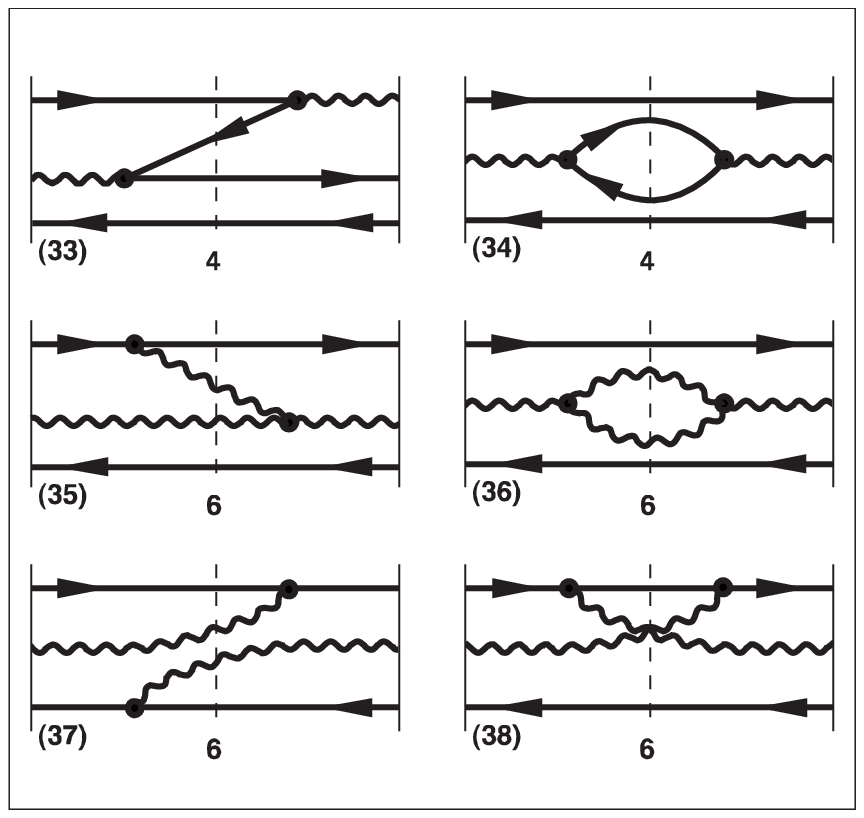,width=50mm} 
\vfill
\caption{\label{fig:6_3}
    Some six graphs of the  participant interaction 
    in the $q\bar q\,g$-space.
}\end{minipage}
\end{figure}

Let us discuss in some greater detail the structure of the
sector Hamiltonians, particularly of $H_{q\bar q\,g}$ 
in Eq.(\ref{eq:620}).
The corresponding graphs are displayed 
diagrammatically in Figs.~\ref{fig:6_2}  and \ref{fig:6_3}. 
Those in Fig.~\ref{fig:6_2} differ from look those in 
Fig.~\ref{fig:6_1} only by an additional gluon.
The gluon does not change quantum numbers under impact
of the interaction and acts like a spectator.  
The graphs in Fig.~\ref{fig:6_2} will be referred to as the  
`spectator interaction' $\overline U _3$. Correspondingly, 
the graphs of Fig.~\ref{fig:6_3} will be referred to
as the `participant interaction' $\widetilde U _3$.  
The gluon quantum numbers are changed and the gluon
participates in the interaction. The analogous separation 
into spectator and participant  interactions
can be made in all quark-pair-glue sector Hamiltonians:
 \begin {equation}  
     H _n =  T  _n + \overline U _n + \widetilde U _n
\ , \quad{\rm for}\quad  n= 3,6,10,15,\dots
\ .  \label {eq:626} \end {equation} 
Note that the spectator interaction has the same diagrams 
as in Fig.~\ref{fig:6_1},  except for the additional 
{\em free and non-interacting gluons}.

The essence is that the spectator interactions $\overline U _n$
define mean fields, and that they
can be associated with their own resolvents $\overline G _n$:
\begin{equation} 
     \overline G _n  =   {1 \over  \omega - T _n - \overline U _n }
\,, \label{eq:641}\end{equation}  
while $G_n$ was  
\begin{equation} 
G _n   =
     {1 \over  \omega - T _n - \overline U _n - \widetilde U _n }
\,. \end{equation}  
Both are related by
$     G _n    =   \overline G _n   +  \overline G _n \,
                        \widetilde U _n \, G _n $.
Equivalently, they can be written as an infinite series
$     G _n  = \overline G _n
     + \overline G _n \widetilde U _n\, \overline G _n  
     + \overline G _n \widetilde U _n\, \overline G _n 
                                     \widetilde U _n\, \overline G _n   
     + \dots\ $,
as usual. The main difference to the above conventional  
series is, that there the 
`unperturbed propagator' $ G _0 (\omega) $ 
refers to the system without interactions while here the 
`unperturbed propagators' $\overline G _n$ include 
the mean fields $\overline U_n$.
One therefore deals here with `perturbation theory in medium'.
Note that the present series differ from the above also 
with respect to the physics: 
The system stays in sector $n$. 
This allows for  the identical rearrangement 
\begin{eqnarray} 
   G _n &=& R_n\,\overline G _n \,R_n
\,, \label {eq:6520}\\{\rm with}\quad
  R_n &=& 1
  +  {1\over 2} \overline G _n \widetilde U _n
  +  {3\over 8} \overline G _n \widetilde U _n   \,
                           \overline G _n \widetilde U _n + \dots  
\nonumber\\
  &=&  {1\over \sqrt { 1 - \overline G _n \widetilde U _n}  }
\,.\label {eq:6510}\end{eqnarray} 
This can be verified order by order and has, 
to our recollection, not been given before.
The square matrices $R_n$ are always sandwiched between a 
quark-pair-glue resolvent $\overline G$ and the vertex $V$.
It is therefore meaningful to introduce the  effective vertex
$\overline V = R_nV$ or $\overline V = R_nV$, thus
\begin  {equation} 
     V\,G _n \, V = \overline V \,\overline G_n\,\overline V      
\ . \label {eq:652} \end {equation} 
The vertex operators $R_n$ come with same power 
as the coupling constant. 
They are a genuine part of the interaction and 
accumulate the many-body aspects manifesting themselves 
as non-point-like vertices. 
Finally, rewriting systematically
Eqs.(\ref{eq:610})-(\ref{eq:622}), one gets
\begin {eqnarray} 
     \overline H _6 
&=&   T_6 + \overline V \,\overline G _ {10} \overline V
                     + \overline V \,\overline G _{10} \overline V  
              G _9 \overline V \,\overline G _ {10} \overline V 
\ , \label {eq:662} \\
     \overline H _3 
&=&  T_3 + \overline V \,\overline G _6 \overline V 
                    + \overline V \,\overline G _6 \overline V  
              G _5\overline V \,\overline G _6 \overline V 
\ , \label {eq:663} \\      
       H _1= \overline H _1
&=&  T_1 + \overline V \,\overline G _3 \overline V 
                    + \overline V \,\overline G _3 \overline V  
             G _2 \overline V \,\overline G _3 \overline V 
\ . \label {eq:664} \end {eqnarray}
Instead of being similar, the quark-pair-glue sector
Hamiltonians  $\overline H _n = T_n + \overline U _n$ 
now apparently all look the same. 
Note that the resolvents 
$G _2$, $G _5$ and $G _9$ carry no bar. They correspond to
the pure glue sectors. The distinction between participants
and spectators makes no sense there.

\section{The approximations}

Thus far the approach is formally exact. For calculating 
the effective interaction the two resolvents 
$\overline G_3$ and $\overline G_2$ and the vertex function
$R_3$ are needed. How can they be approximated?
Suppose to have solved the eigenvalue problem in the 
$q\bar q$-space, 
\begin  {equation}
    \sum _{q^\prime,\bar q^\prime}
    \langle q;\bar q\vert H_{q\bar q} (\omega)
    \vert q^\prime;\bar q^\prime\rangle
    \psi_b(\omega) \rangle =  M_{b} ^2(\omega) 
    \langle q;\bar q\vert\psi_b(\omega) \rangle
\,.\label{eq:6.60}\end {equation} 
The eigenvalues are enumerated by
$b=1,2,\dots$. The  corresponding eigenfunctions 
$\langle q;\bar q\vert\psi_b(\omega) \rangle$ are a complete set.  
Despite working in the continuum limit, we continue to use 
summation symbols for the sake of a more compact notation. 
Suppose further that an $\omega$ was found 
which has the same value as the lowest eigenvalue $M^2=M^2_1$. 
The substitution $\omega=M^2$ will hence forward be done
without explicitly mentioning.
Next, ask for the eigenvalues and eigenfunctions
in the $q\bar q\,g$-space. One need not to solve 
another eigenvalue problem,  
since the result is known ahead of time!
By construction, the gluon moves
relative to the meson subject to momentum conservation.
The eigenfunction is a thus a product state 
$\vert \psi_{b,s}\rangle
   =\vert \psi_{b}\rangle
   \otimes \vert \varphi_{s}\rangle$. 
Parameterizing the  gluons four-momentum as 
$    p_g^\mu = (yP^+, y\vec P_{\!\perp} + \vec q_{\!\perp},
    p_g^-) $, 
the eigenvalues are
\begin{equation}
    M_{b,s} ^2
    = {M_b^2 + \vec q_{\!\perp}^{\,2} \over (1-y)} 
    + {\vec q_{\!\perp}^{\,2} \over y} 
\,.\end{equation}
Knowing the eigenvalues and eigenfunctions, one can
calculate the exact resolvent. After a few identical rewritings
one gets
\begin{equation}
      \langle q;\bar q;g \vert \overline G _3
      \vert q ^\prime;\bar q^\prime;g ^\prime \rangle =
      \overline G _3 (q;\bar q;g)\left[
      \langle q;\bar q;g            \vert q ^\prime;\bar q^\prime;g^\prime\rangle -
      \langle q;\bar q;g\vert A\vert q ^\prime;\bar q^\prime;g^\prime\rangle 
      \right]
\,. \label{eq:6.69}\end{equation}
The operator $A$ can never become a Dirac-$\delta$ function 
$\langle q;\bar q;g\vert q^\prime;\bar q^\prime;g^\prime\rangle$,
since
\begin{equation}
    A= \sum _{b,s} \displaystyle
    \vert\psi_{b,s}\rangle\,{\displaystyle y(M_b^2-M^2)
    \over  \displaystyle Q^2+y(M_b^2-M^2)}\, 
    \langle \psi_{b,s} \vert 
\,.\end{equation}
It is therefore dropped, as an approximation. 
The proportionality coefficient is 
\begin{equation}
    \overline G _3 (q;\bar q;g)
    = -{y(1-y) \over Q^2}
\ , \quad{\rm with}\quad
    Q^2 = y^2 M ^2 +\vec q_{\!\perp}^{\,2}
\,.\label{eq:6.68}\end{equation}
The same approximation yields
for the two gluon propagator
\begin{equation}
   \overline G _2 (g_1;g_2)={1\over M^2 - M^2_{gb} }
\,.\end{equation}
It is  parameterized 
in terms of the glue ball mass $M^2_{gb}$.

For calculating the operator $R_3^2$, 
one addresses first to calculate the operator
$B:=-\overline G_3(V\overline G_4V +V\overline G _6V)$. 
An inspection of Fig.~\ref{fig:6_2} yields that the loop diagrams
(34) and (36) are the leading terms, by far.
The have been calculated long ago by 
Thorn \cite{tho79} for $n_c=3$ colors and massless quarks,
using the perturbative propagators. Repeating this calculations
with the non-perturbative propagators 
$\overline G_4$ and $\overline G_6$, which can be approximated 
in line with Eq.(\ref{eq:6.69}), one gets a diagonal $B$ with
\begin{equation}
    B= - {1\over4}\alpha_s\,c_0\,
    \ln{\left[1+{1\over(1-y)} {\Lambda^2\over Q^2}\right]}
\,,\label{eq:6.79}\end{equation}
with $\alpha_s={g^2/4\pi}$ and 
$c_0 \equiv (33-2N_f){/6\pi}$. The latter number is not unfamiliar 
from the work of Gross and Wilczek \cite{grw73}.
Details will be given elsewhere.
Applying the same procedure to even higher spaces gives
again and again the same type of vacuum polarization diagrams,
and thus a diagonal
\begin{equation}
    R_3^2 = 
    {1\over\displaystyle 1+
    {B\over\displaystyle 1+
    {B\over\displaystyle 1+
    \dots }}}
\,.\end{equation}
This continued fraction can be resumed to all orders
by $R_3^2= {1/(1+BR_3^2)}$ and solved
\begin{equation}
     R_3^2 = {2\over 1+\sqrt{ 1+ \alpha_s c_0 \,
     \ln{\left[1+{1\over(1-y)} {\Lambda^2\over  Q^2}\right]} } }
\,.\label{eq:6.87}\end{equation}
Now, one can calculate analytically
the effective interaction!

\section{Outlook}
The strength of the interaction ($R_3$) tends to zero for increasing
cut-off $\Lambda\rightarrow\infty$.
One may (or may not) interpret this result as 
{asymptotic freedom}.  But this would be premature.
The above is only the {\em regulated effective interaction},
albeit being well defined for any fixed value of the cut-off 
$\Lambda$. The dependence on $\Lambda$ must be removed
by a future {renormalization group analysis} \cite{wwh94}.
It looks as if the present explicit form is almost ideally suited 
for an application of the Hamiltonian flow equations 
\cite{weg94}. Finally, one should emphasize that the
above approach is based on comparatively weak approximations
being thoroughly non-perturbative: No smallness
assumption on the coupling constant was ever needed,
nor was the Fock space ever truncated. Gauge invariance is
maintained, and Lorentz covariance (including rotations)
is strictly observed in the continuum limit.

\end{document}